\documentstyle[prl,preprint,aps,epsf]{revtex}

\newcommand{\postscript}[2]{
\setlength{\epsfxsize}{#2}
\centerline{\epsfbox{#1}}}           
\newlength{\headroom}
\newlength{\psfigskip}
\begin{document}
\title{
		     Cosmological Implications of Two 
		     Conflicting Deuterium Abundances
}
\author{             
		     Naoya Hata and Gary Steigman, 
}
\address{
{\it
		     Department of Physics,
		     The Ohio State University,               \\
	             Columbus, Ohio 43210                     \\
}}
\author{             Sidney Bludman and Paul Langacker        }
\address{
{\it
                     Department of Physics,                       
                     University of Pennsylvania,               \\              
                     Philadelphia, Pennsylvania 19104          \\ 
}}

                     \date{ March 16, 1996, 
                            OSU-TA-6/96} 

\maketitle
%

\renewcommand{\baselinestretch}{0.95}
\begin{abstract}

Constraints on big bang nucleosynthesis (BBN) and on cosmological
parameters from conflicting deuterium observations in different high
red-shift QSO systems are discussed.  The high deuterium observations
by Carswell {\it et al}., Songaila {\it et al}., and Rugers \& Hogan
is consistent with $^4$He and $^7$Li observations and Standard BBN
($N_\nu$ =3) and allows $N_\nu \leq 3.6$ at 95\% C.L., but is
inconsistent with local observations of D and $^3$He in the context of
conventional theories of stellar and Galactic evolution.  In contrast,
the low deuterium observations by Tytler, Fan \& Burles and Burles \&
Tytler are consistent with the constraints from local Galactic
observations, but require $N_\nu = 1.9 \pm 0.3$ at 68\% C.L.,
excluding Standard BBN at 99.9\% C.L., unless the systematic
uncertainties in the $^4$He observations have been underestimated by a
large amount.  The high and low primordial deuterium abundances imply,
respectively, $\Omega_{\rm B}h^2 = 0.005 - 0.01$ and $\Omega_{\rm
B}h^2 = 0.02 - 0.03$ at 95\% C.L.  When combined with the high baryon
fraction inferred from x-ray observations of rich clusters, the
corresponding total mass densities (for $50 \le H_0 \le 90$) are
$\Omega_{\rm M} = 0.05 - 0.20$ and $\Omega_{\rm M} = 0.2 - 0.7$,
respectively (95\% C.L.)  The range of $\Omega_{\rm M}$ corresponding
to high D is in conflict with dynamical constraints ($\Omega_{\rm M}
\ge 0.2 - 0.3$) and with the shape parameter constraint 
($\Gamma = \Omega_{\rm M}h = 0.25 \pm 0.05$) from large scale
structure formation in CDM and $\Lambda$CDM models.

\end{abstract}
\pacs{PACS numbers: }
%

\renewcommand{\baselinestretch}{1.00}

\section{Introduction}

Among the light nuclides synthesized during the early evolution of the
universe, deuterium is unique in its sensitivity to the universal
density of baryons and in the simplicity of its galactic evolution.
As gas is incorporated into stars and the heavy elements are
synthesized, D is only destroyed \cite{Epstein-Lattimer-Schramm}
so that any D abundance inferred from observations provides a
{\em lower} bound to its primordial value.  
\footnote{
$X_{2{\rm P}} > X_{2{\rm OBS}}$, where the D mass fraction is $X_2 = 2
X\; n_{\rm D}/n_{\rm H}$; $X$ is the hydrogen mass fraction, and $n_x$
is the number density for nuclide $x$; the subscript P is for the
primordial abundance.  As an estimate of the primordial value $X_{\rm
P} = 1 - Y_{\rm P}$, we will adopt $X_{\rm P} = 0.76 \pm 0.01$.  In
this paper we quote 1$\sigma$ uncertainties unless otherwise
indicated. }
Unfortunately, an {\em upper} bound to the primordial D abundance is
more uncertain, depending on the evolutionary history of the matter
being observed.  Thus, although estimates of the D abundance in the
presolar nebula \cite{Geiss,Steigman-Tosi-95} and in the local
interstellar medium (ISM) \cite{McCullegh,Linsky-etal} provide
interesting lower bounds to primordial D [$X_{2{\rm P}} \ge
X_{2\odot}= (3.6 \pm 1.3) \times 10^{-5}$, $X_{2{\rm P}} \ge X_{2{\rm
ISM}}= (2.2 \pm 0.3) \times 10^{-5}$, where the H mass fraction has
been taken to be $X_\odot = X_{\rm ISM} = 0.70 \pm 0.01$],
upper bounds are more model dependent (see, for example,
Ref.~\cite{Vangioni-Flam-Audouze,Tosi,Steigman-Tosi-92,Steigman-Tosi-95,%
D-paper}).  For this reason, observations of D in (nearly) unevolved
systems (high red-shift, low metallicity QSO absorbers) have been
eagerly anticipated.  If, indeed, $X_{2{\rm P}} \sim X_{2{\rm QSO}}$,
then because of the sensitivity of the D abundance to the nucleon
abundance ($\eta = n_{\rm B}/n_\gamma$; the ratio of the present
baryon density to the critical density is $\Omega_{\rm B}h^2 = 0.0037
\eta_{10}$, where the Hubble parameter is $H_0 = 100 h$ km/s/Mpc and
$\eta_{10} = 10^{10}\eta$), a measurement of (D/H)$_{\rm QSO}$ to
$\sim$30\% accuracy will lead to a determination of $\eta$ to $\sim$
20\% accuracy.  Armed with $\eta$, reasonably accurate predictions of
the primordial abundances of $^3$He, $^4$He, and $^7$Li will follow
(see, for example, Ref.~\cite{D-paper}).  For example, for $1.5 <
\eta_{10} < 10$, a 20\% uncertainty in $\eta_{10}$ will lead to an
uncertainty in the predicted $^4$He mass fraction which ranges from
$\sim 0.003$ (at low $\eta_{10}$) to $\sim 0.002$ (at high
$\eta_{10}$).  Deuterium is the ideal baryometer.

In the last two years, observations of D in high red-shift, low
metallicity QSO absorbers have begun to appear in the published
literature
\cite{Carswell-etal,Songaila-etal,Rugers-Hogan,Tytler-Fan-Burles,%
Burles-Tytler}.
The first observations of D in absorption against Q0014+813
\cite{Carswell-etal,Songaila-etal,Rugers-Hogan} suggested a surprisingly
high abundance [for our quantitative comparisons we will adopt the
recent reanalysis by Rugers and Hogan: D/H = $(1.9 \pm 0.4) \times
10^{-4}$, $X_2 = (2.9 \pm 0.6) \times 10^{-4}$], roughly an order of
magnitude larger than the presolar or ISM values ($X_{2{\rm
QSO}}/X_{2{\rm ISM}} \sim 8 \pm 3$, $X_{2{\rm QSO}}/X_{2\odot} \sim 13
\pm 3$).  As such efficient D destruction in the Galaxy is not
expected
\cite{Vangioni-Flam-Audouze,Tosi,Steigman-Tosi-92,Steigman-Tosi-95,Edmunds}
it has been suggested that the feature identified as D in Q0014+813
might be a hydrogen interloper \cite{Steigman-94}.  However, the
Rugers-Hogan reanalysis argues against this possibility.  Further,
recent papers \cite{Carswell-etal-1996,Wampler-etal} present evidence
for D absorption in front of two other QSOs (Q0420-388 and
BR1202-0725, respectively) which, if the identifications are correct,
suggest ${\rm D/H} \ge 2 \times 10^{-5}$ and ${\rm D/H} \le 1.5 \times
10^{-4}$, respectively.  Although puzzling from the point of view of
chemical evolution in the Galaxy, the high D abundance points towards
a low baryon density ($\eta_{10} \sim 2$) which is consistent with the
predicted and observed (inferred) primordial abundances of $^4$He and
$^7$Li \cite{Steigman-94,crisis-paper}.  As we shall see, however,
this low baryon density ($\Omega_{\rm B} h^2 \sim 0.007$) is in
conflict with determinations of the total mass density and the baryon
fraction inferred from x-ray observations of rich clusters.

In contrast, from recent observations, Tytler, Fan, and Burles
\cite{Tytler-Fan-Burles} and Burles and Tytler \cite{Burles-Tytler}
derive a low D abundance: $({\rm D/H}) = [2.3 \pm 0.3$ (stat) $\pm
0.3$ (sys) $]\times 10^{-5}$ towards the QSO1937-1009
\cite{Tytler-Fan-Burles} and $({\rm D/H}) = [2.5^{+0.5}_{-0.4}$ (stat)
$^{+0.4}_{-0.3}$ (sys)] $\times 10^{-5}$ towards the QSO1009+2956
\cite{Burles-Tytler}.  We have combined their two results to obtain
(D/H)$_{\rm QSO} = (2.4 \pm 0.5) \times 10^{-5}$; $X_{2{\rm QSO}} =
(3.6 \pm 0.8) \times 10^{-5}$.  Although marginally larger than ISM
deuterium ($X_{2{\rm QSO}}/X_{2{\rm ISM}} = 1.6 \pm 0.4$), the low D
abundance is not very different from the presolar value ($X_{2{\rm
QSO}}/X_{2\odot} = 1.0 \pm 0.4$), suggesting that even though the
absorbers are at high redshift ($z_{\rm abs}$ = 3.572 and 2.504) and
have very low metallicity ($\sim 10^{-3}$ solar), some D may have
already been destroyed ($X_{2{\rm P}} \ge X_{2{\rm QSO}}$).  If indeed
$X_{2{\rm P}} \sim X_{2{\rm QSO}}$ (no significant D destruction),
then the problems for BBN identified by Hata {\it et al}.\
\cite{crisis-paper}, which were based on $X_{2{\rm P}}$ inferred from
solar system observations of D and $^3$He, persist.  The higher baryon
density suggested by the low D result is, however, in agreement with
the x-ray cluster data (but still supports a low density universe).

It is hoped that future observations of D in other high red-shift, low
metallicity QSO absorbers will resolve the current dichotomy between
the high D result for Q0014+813 on the one hand
\cite{Carswell-etal,Songaila-etal,Rugers-Hogan} and the 
low D results for Q1937-1009 and Q1009+2956 on the other
\cite{Tytler-Fan-Burles,Burles-Tytler}.  Here, we
explore the implications for cosmology (the baryon density), for the
primordial abundances of the other light nuclides ($^4$He and $^7$Li),
and for particle physics (bounds to the effective number of equivalent
light neutrinos, $N_\nu$), of the high D abundance and contrast them
with those for the low D abundance.

\section{Primordial D and BBN}

If (D/H)$_{\rm P}$ is fixed, standard big bang nucleosynthesis (SBBN:
homogeneous, $N_\nu = 3$, the neutron life time $\tau_n = 887 \pm 2$
s, etc) can be used to predict the primordial abundances of the other
light nuclides and determine the present baryon density.  In
Fig.~\ref{fig:bbn_comb_qsos} the SBBN predicted abundances of $^4$He
(mass fraction $Y$), D ($y_2$ = D/H), and $^7$Li ($y_7$ = $^7$Li/H) are
shown as a function of the nucleon to photon ratio $\eta$ for $1 \le
\eta_{10} \le 10$.   Convolving the SBBN predictions (including
uncertainties estimated by the Monte Carlo method of
Ref~\cite{SBBN-analysis}) with the high D and low D results constrains
$\eta$ and leads to predictions of $Y$ and $y_7$ as may be seen in
Fig.~\ref{fig:bbn_comb_qsos}.  Also shown in
Fig.~\ref{fig:bbn_comb_qsos} are the 68 and 95\% C.L.\ contours for
the overlap between the inferred primordial abundances of $^4$He
[$Y_{\rm P} = 0.232 \pm 0.003$ (stat) $\pm 0.005$ (sys)] and $^7$Li
[log $y_7 = -9.8 \pm 0.2$ (sys) $\pm 0.3$ (depletion/creation)]
\footnote{ %
The statistical uncertainty in $^4$He is assumed to be Gaussian, while
the systematic uncertainties in $^4$He and $^7$Li and the uncertainty
in $^7$Li depletion/creation are treated as flat (top hat)
distributions.  The statistical uncertainty in $y_7$ is small compared
to the systematic and depletion/creation uncertainties. }
and the BBN predictions.  When using the low D value we must ensure
that consistency with the ISM (and solar system) value is maintained
($X_{2{\rm P}} \ge X_{2{\rm ISM}}$; $X_{2{\rm P}} \ge X_{2\odot}$).
In Fig.~\ref{fig:y2-P_comb} we show the SBBN likelihood distribution
(solid curve) for the low D result [$\log y_2^{\rm QSO} = -4.62 \pm
0.05 \pm 0.06$; $y_2^{\rm QSO} = (2.4 \pm 0.5) \times 10^{-5}$].
Requiring that the QSO D abundance be no smaller than the ISM D
abundance [$y_2^{\rm QSO} \ge y_2^{\rm ISM} = (1.6 \pm 0.2) \times
10^{-5}$], modifies the distribution to the dotted curve in
Fig.~\ref{fig:y2-P_comb}, very slightly truncating the lower end of
the $y_2$ distribution [$y_2^{\rm QSO/ISM} = (2.4 \pm 0.5) \times
10^{-5}$].  It is this latter distribution which we will use in our
comparisons.  If, further, we also require that the QSO D abundance
exceeds that inferred for the presolar nebula [$y_{2\odot} = (2.6 \pm
0.9) \times 10^{-5}$] \cite{Geiss}, the distribution (dashed curve in
Fig.~\ref{fig:y2-P_comb}) is slightly shifted to higher values
[$y_2^{{\rm QSO/ISM/}\odot} = (2.6 \pm 0.5) \times 10^{-5}$].  Given
that the solar system data may be subject to different systematic
errors than those associated with the QSO and ISM absorption
observations, we will limit our analysis to those which follow from
the marginally less restrictive QSO/ISM constraint.  The resulting
SBBN constraints on $\eta$, $\Omega_{\rm B}h^2$, $N_\nu$, $Y$, and
$y_7$ which follow from the high and low-QSO/ISM D abundances are
summarized in Table~\ref{tab:constraints} along with, for comparison,
the previous Hata {\it et al}.\ results \cite{D-paper} which utilized
solar system D and $^3$He abundances.  Fig.~\ref{fig:eta-P_qsos} shows
the likelihood distributions for $\eta$ for high and low primordial
deuterium.

A glance at Fig.~\ref{fig:bbn_comb_qsos} reveals the well-known result
\cite{Dar} that high-D is consistent with the primordial $^4$He abundance 
inferred from HII region data.  In Fig.~\ref{fig:eta-Y_qsos} is shown
the SBBN predicted $^4$He mass fractions corresponding to the two
deuterium values.  It is clear from Figs.~\ref{fig:bbn_comb_qsos} and
\ref{fig:eta-Y_qsos} that, in the absence of large systematic errors
in $^4$He, the observed $^4$He abundance favors high-D and is
inconsistent with low-D.

In Fig.~\ref{fig:eta-y7_qsos} we compare the SBBN predictions of
$^7$Li for high and low deuterium with that inferred from observations
of the Pop II halo stars
\cite{Spite-Spite,Thorburn,Vauclair-Charbonnel,Molaro-Primas-Bonifacio}.
As is clear from Figs.~\ref{fig:bbn_comb_qsos} and
\ref{fig:eta-y7_qsos}, and Table~\ref{tab:constraints}, consistency
with lithium is achieved for both high and low D.  Notice, however,
that while the high-D (low $\eta$) overlap with SBBN bounds stellar
destruction/dilution of $^7$Li ($y_7 \le 3.0 \times 10^{-10}$; log
$y_7 \le -9.5$), the low-D (high $\eta$) overlap actually requires
some modest destruction/dilution ($3.0 \le 10^{10} y_7 \le 7.8$; $-9.5
\le$ log $y_7 \le -9.1$).

Finally, we turn to $^3$He, whose post-BBN evolution is model
(galactic chemical evolution) dependent.  However, since production of
$^3$He by low mass stars can only increase the $^3$He abundance,
observations of D and $^3$He constrain the primordial D and $^3$He
abundances \cite{YTSSO,WSSOK,Steigman-Tosi-95,D-paper}.  For any
chemical evolution history the observed and primordial abundances of D
and $^3$He may be related through one parameter $g_3$, the effective
$^3$He stellar survival fraction, which contains all the stellar and
galactic evolution uncertainties
\cite{YTSSO,Steigman-Tosi-95,D-paper}.  While for a single generation
of stars $g_3 \ge 0.25$ 
\cite{Dearborn-Schramm-Steigman,Dearborn-Steigman-Tosi}, many
specific evolution models suggest $g_3 \ge 0.5$ 
\cite{Steigman-Tosi-92,Palla-etal,Dearborn-Steigman-Tosi}.  Following
Ref.~\cite{D-paper} we show in Fig.~\ref{fig:y3-y2_qsos} the allowed
regions in the $y_{3{\rm P}}$--$y_{2{\rm P}}$ plane inferred from
SBBN and the high and low deuterium abundances.  Although the low-D
data is entirely consistent with the galactic evolution of D and
$^3$He, the high-D data requires a surprisingly small value of $g_3$
($\le 0.10$ at 95\% C.L.) for consistency.  Indeed, the high-D data
suggests that more than 90\% of the present ISM has been cycled
through stars (since D would have to have been destroyed by a factor
of $\sim 10$).  With such efficient processing of gas through stars,
the low metallicity of the ISM is a challenge to galactic chemical
evolution models \cite{Vangioni-Flam-Audouze,Tosi}.

\section{SBBN and $N_\nu$}

  From the discussion above it is clear that high primordial D is
entirely consistent with the predictions of SBBN and the observed
abundances of $^4$He and $^7$Li.  For low primordial D there is a
significant tension between the predictions of SBBN and the inferred
primordial abundance of $^4$He.  If we allow $N_\nu$, the equivalent
number of light neutrinos, to depart from the SBBN value $N_\nu = 3$,
we may use the combined D, $^4$He, and $^7$Li data to find the best
$N_\nu$.  Fig.~\ref{fig:nnu-P_comb} shows the $N_\nu$ likelihood
distributions for high and low (QSO/ISM) D; for comparison we also
show the distribution derived from solar system D and $^3$He with $g_3
= 0.25 - 0.50$ \cite{D-paper}.  Of course, it is always possible that
$Y_{\rm P}$, inferred from nearly primordial (low metallicity)
extragalactic HII regions
\cite{Pagel-etal,Skillman-Kennicutt,Olive-Steigman-He4}, has
been underestimated due to systematic errors.  Although such
uncertainties (ionization corrections, collisional excitation
corrections, corrections for dust, corrections for stellar absorption,
etc.) could either decrease or increase the inferred value ($Y_{\rm P}
= 0.232 \pm 0.003$), recent work has emphasized those corrections
which might increase $Y_{\rm P}$
\cite{Sasselov-Goldwirth,Copi-Schramm-Turner-I}.  If we write
$Y_{\rm P} = 0.232 \pm 0.003 + \Delta Y_{\rm sys}$ (where $\Delta
Y_{\rm sys}$ $\ge$ 0), then there is a direct relation between $\Delta
Y_{\rm sys}$ and $N_\nu$, which we show for high and low deuterium in
Fig.~\ref{fig:dY-nnu_qsos}.  For $\Delta Y_{\rm sys} \le 0.009$,
high-D and SBBN are consistent ($N_\nu = 3$), but, if $\Delta Y_{\rm
sys}$ is larger, $N_\nu \ge 3$ would be required.  In contrast, SBBN
and low-D are inconsistent unless $\Delta Y_{\rm sys}
\ge 0.011$.

\section{$\Omega_{\rm B}$ and $\Omega_{\rm M}$}

Either choice of high or low deuterium leads, through SBBN,
to reasonably tight constraints on the present ratio of nucleons to
photons (see Fig.~\ref{fig:bbn_comb_qsos} and
Table~\ref{tab:constraints}), thus bounding the present universal
density of baryons $\rho_{\rm B}$.  Comparing $\rho_{\rm B}$ to the
critical density $\rho_c$, we have
\begin{equation}
	\Omega_{\rm B} = {\rho_{\rm B} \over \rho_c}
                       = 3.66 \times 10^{-3} \eta_{10} h^{-2}.
\end{equation}
In Fig.~\ref{fig:H-Omega_B} we show the $\Omega_{\rm B}$ vs. $H_0$
relation, where the two bands correspond to the 68 and 95\% C.L.\
ranges for $\eta_{10}$ allowed by the QSO D abundances (see
Table~\ref{tab:constraints}).  Also shown in Fig.~\ref{fig:H-Omega_B}
is an estimate of the luminous baryons identified by observations
in the radio, optical, ultra-violet, and x-ray parts of the
spectrum\cite{Persic-Salucci},
\begin{equation}
	\Omega_{\rm LUM} = 0.004 + 0.0007 h^{-3/2}.
\end{equation}
Over the entire range of $H_0$, $\Omega_{\rm B} \ge \Omega_{\rm LUM}$,
suggesting the presence of dark baryons.  As a minimum estimate of the
total density (baryons plus non-baryonic dark matter) inferred from
the dynamics of groups, clusters, etc., we have adopted $\Omega_{\rm
DYN} \ge 0.2$
\cite{Ostriker-Steinhardt,Shaya-Peebles-Tully,Carlberg-Yee-Ellingson}
(although others have suggested $\Omega_{\rm DYN} \gtrsim 0.3$
\cite{Dekel-Rees}).  As can be seen from Fig.~\ref{fig:H-Omega_B}, 
unless $H_0$ is very small (and D is very low), $\Omega_{\rm DYN} \ge
\Omega_{\rm B}$, providing support for the presence of non-baryonic
dark matter.

X-ray emission from the hot (baryonic) gas in rich clusters of
galaxies offers a valuable probe of the fraction of the total mass in
the universe contributed by baryons.  Although relatively rare, such
large mass concentrations are expected to provide a fair sample of
$f_{\rm B} = \Omega_{\rm B}/\Omega_{\rm M}$, where $\Omega_{\rm M}$ is
the total matter density parameter
\cite{White-Frenk,White-Fabian,Evrard-Metzler-Navarro}.  For clusters,
$f_{\rm B} = M_{\rm B} / M_{\rm TOT} > M_{\rm HG}/M_{\rm TOT}$, where
$M_{\rm HG}$ is the mass of the x-ray emitting hot gas in the cluster
and $M_{\rm TOT}$ is the total mass which determines the cluster
binding.  It is conventional to write $f_{\rm HG} = f_{50}
h_{50}^{-3/2}$, where $h_{50} = H_0 / 50$ km/s/Mpc, so that
\begin{equation}
	\Omega_{M} h_{50}^{1/2} < 0.0146 \eta_{10}/f_{50}.
\label{Eqn:x-ray-constraint}
\end{equation}
The inequality in (3) arises from the neglect of the baryons in the
galaxies of the cluster; their inclusion would reduce the upper bound
on $\Omega_{M}$ by $\lesssim$ 5--20\% (depending on $H_0$).  However,
since the presence of other (dark) baryons (e.g., Machos) cannot be
excluded observationally and may be large \cite{Gould}, the cluster
data is best utilized to provide an $\it{upper}$ bound to
$\Omega_{M}$.  Thus, in our subsequent analysis we shall employ x-ray
data and BBN to evaluate the right hand side of
Eqn.~\ref{Eqn:x-ray-constraint} which we will use to provide an upper
bound to $\Omega_{\rm M}$ (as a function of $H_0$).

The surprise provided by x-ray observations of rich clusters has been
the relatively large baryon fraction ($f_{50} \ge 0.1 - 0.2$) which,
when coupled to the relatively low upper bound on $\eta_{10}$ from BBN,
has led to the ``X-Ray Cluster Baryon Catastrophe''
\cite{White-Frenk,White-Fabian,Steigman-Felten}: 
$\Omega_{\rm M} < 1$ unless $H_0$ is very small.

Following the recent analysis of Evrard {\it et al.}
\cite{Evrard-Metzler-Navarro} and Evrard \cite{Evrard}, we adopt
$f_{50} = 0.20 \pm 0.03$, and use this and the bounds on $\eta_{10}$
 from high and low D (see Table~\ref{tab:constraints}) to constrain the
$\Omega_{M}$ vs.\ $H_0$ relation (Eqn.~\ref{Eqn:x-ray-constraint}) in
Fig.~\ref{fig:H-Omega_M}.  We have allowed $H_0$ to remain
unconstrained although we believe that recent data suggest $H_0 = 70
\pm 15$ km/s/Mpc ($h = 0.7 \pm 0.15$ and $h_{50} = 1.4 \pm 0.3$).
The x-ray cluster constraints require low $\Omega_{M}$, excluding the
preferred Einstein-de Sitter value of $\Omega_{\rm M} = 1$ unless
$f_{50}$ and/or $H_0$ is much smaller than data indicate.

Further evidence for low $\Omega_{\rm M}$ in the context of cold dark
matter models (CDM) comes from large scale structure constraints on
the shape parameter (see, for example, \cite{Peacock-Dodds}): $\Gamma
= \Omega_{\rm M}h = 0.25 \pm 0.05$.  Since the popular inflationary
paradigm suggests that the 3-space curvature may vanish, evidence in
favor of low $\Omega_{\rm M}$ has led to consideration of an
alternative cosmology with a non-vanishing cosmological constant
($\Lambda$) such that $\Omega_{\rm TOT} = \Omega_{\rm M} +
\Omega_\Lambda = 1$ (see, for example,
\cite{Steigman-Felten}).   Such $\Lambda$ cold dark matter models
($\Lambda$CDM) provide the additional benefit of helping to resolve
the ``age problem'' (for the same value of $\Omega_{\rm M}$, and fixed
$H_0$, a $\Lambda$CDM universe is older than the corresponding CDM
universe).  In Figs.~\ref{fig:H-Omega_M_comb} and
\ref{fig:H-Omega_ML_comb} we show (at 68\% C.L.) the regions in
the $H_0$--$\Omega_{\rm M}$ plane consistent with the x-ray/BBN
constraints (for high and low D) and with the shape parameter
($\Gamma$).  Also shown is the $\Omega_{\rm M}$--$H_0$ relation for two
choices of the present age of the universe ($t_0$ = 12 and 15 Gyr)
along with the dynamically inferred lower bound to the mass density
($\Omega_{\rm DYN}$).  In both cases (CDM and $\Lambda$CDM), the
low-D, high-$\eta_{10}$ choice is preferred over the high-D,
low-$\eta_{10}$ result.  

A third popular cosmology is the mixed, hot plus cold dark matter
model (HCDM; see for example,
\cite{Liddle-etal,Klypin-Nolthenius-Primack,Ma-Bertschinger}).  In its
standard version it is assumed that $\Omega_{\rm M} = 1$, but that
$\sim 20 - 30$\% of $\Omega_{\rm M}$ is in hot dark matter (e.g.,
neutrinos with mass of 1 -- 10 eV) which is relatively unclustered on
large scales.  The shape parameter constraint is not relevant to
constraining the HCDM model, but the requirement that $\Omega_{\rm
TOT} = 1$ exacerbates the x-ray cluster baryon catastrophe
\cite{White-Frenk,White-Fabian,Evrard-Metzler-Navarro,Steigman-Felten}
and the age problem.  Even if all HDM could be excluded from x-ray
clusters \cite{Strickland-Schramm}, which seems unlikely
\cite{Kofman-etal}, $\Omega_{\rm M} \gtrsim 1 - \Omega_{\rm HDM}$.
Coupled with the upper bound from the x-ray cluster data, $\Omega_{\rm
M} < 0.7$, this requires $\Omega_{\rm HDM} \gtrsim 0.3$, nearly
closing the preferred window ($0.2 \lesssim \Omega_{\rm HDM} \lesssim
0.3$) on HCDM models.

\section{Discussion}

A determination of the deuterium abundance in a nearly uncontaminated
environment such as that provided by high redshift, low metallicity
QSO absorption clouds could be a key to testing the consistency of
primordial nucleosynthesis in the standard, hot, big bang cosmology,
to pinning down the universal density of baryons, and to constraining
physics beyond the standard model of particle physics.  Such data is
beginning to be acquired but, at present, the observational situation
is in conflict.  On the one hand there is evidence in favor of high D
\cite{Songaila-etal,Carswell-etal,Rugers-Hogan,Wampler-etal,%
Carswell-etal-1996}: (D/H)$ \sim 2 \times 10^{-4}$.  In contrast,
Tytler, Fan, \& Burles \cite{Tytler-Fan-Burles} and Burles \& Tytler
\cite{Burles-Tytler} find evidence for low D: (D/H)$ \sim 2 \times
10^{-5}$.  If the former, high-D values are correct, it is surprising
that Tytler, Fan, \& Burles and Burles \& Tytler fail to find such a
large abundance in their high redshift ($z$ = 3.57 and 2.50), very low
metallicity ($Z/Z_\odot \sim 10^{-3}$) absorbers; high $z$ and low $Z$
argue against an order of magnitude destruction of primordial D.  If,
instead, the low D result is correct, such weak D-absorption might
often go unnoticed and the high-D cases might be accidental
interlopers.  Based on velocity information Rugers and Hogan
\cite{Rugers-Hogan} argue against this possibility which, if more
high-D cases are found, will become increasingly unlikely.
Presumably, the present confused situation will be clarified by the
acquisition of more data.  Here, we have considered separately the
consequences for cosmology and particle physics of the high-D and
low-D data.

\begin{figure}[h]

\postscript{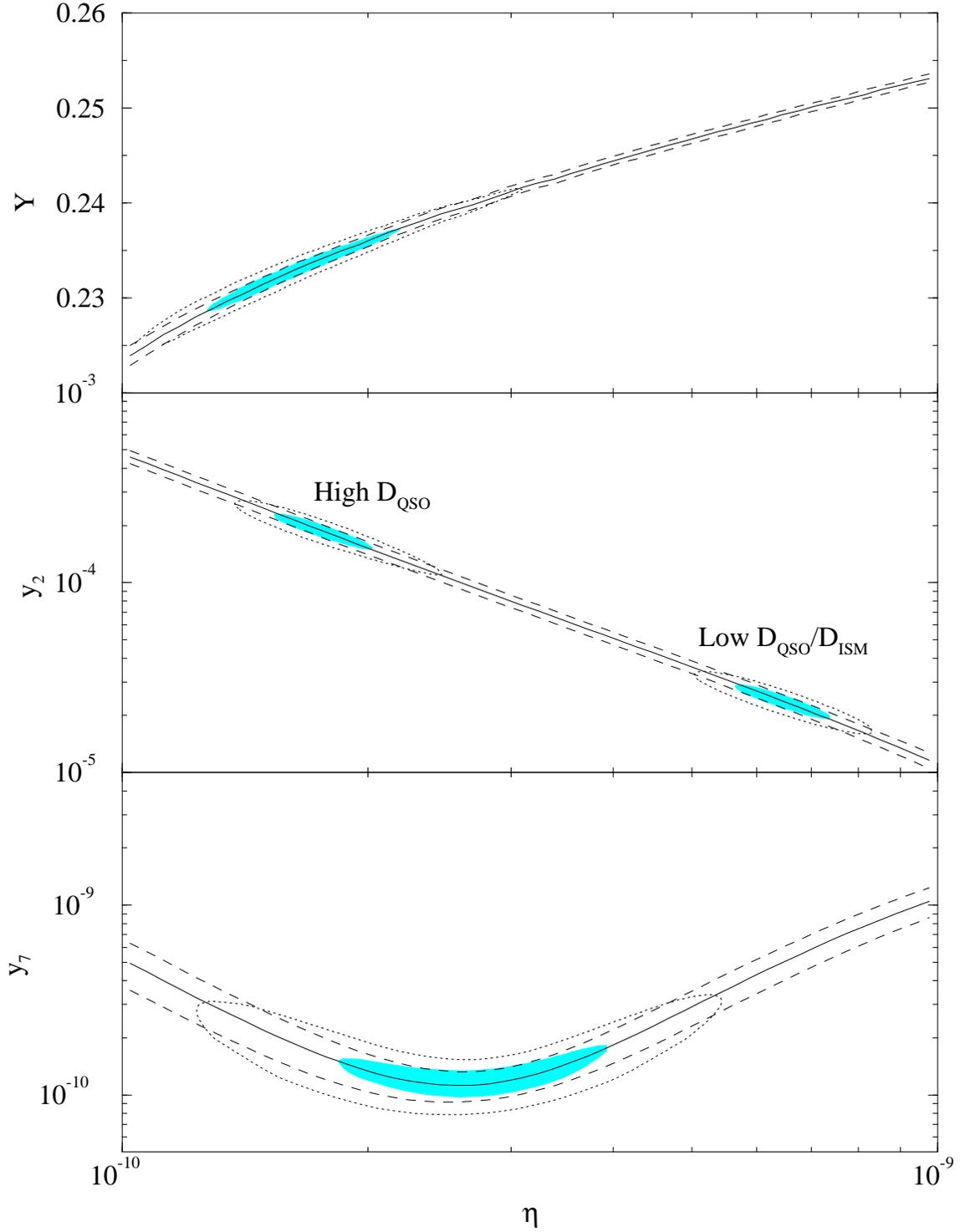}{0.95\hsize}
\vspace*{-3ex}

\caption{
BBN predictions (solid lines) for $^4$He ($Y_{\rm P}$), D ($y_{2{\rm
P}}$), and $^7$Li ($y_{7{\rm P}}$) with the theoretical uncertainties
(1$\sigma$) estimated by the Monte Carlo method (dashed lines).  Also
shown are the regions constrained by the observations at 68\% and 95\%
C.L. (shaded regions and dotted lines, respectively).  We use the QSO
measurements for $y_{2{\rm P}}$ from Ref.~\protect\cite{Rugers-Hogan}
and \protect\cite{Tytler-Fan-Burles,Burles-Tytler}.}
\label{fig:bbn_comb_qsos}
\end{figure}

For the high-D case, SBBN ($N_\nu = 3$) is consistent with the
inferred primordial abundances of D, $^4$He, and $^7$Li provided that
the baryon density is small (see Table~\ref{tab:constraints} and
Figs.~\ref{fig:bbn_comb_qsos}, \ref{fig:eta-P_qsos},
\ref{fig:eta-Y_qsos},
\ref{fig:eta-y7_qsos}, \ref{fig:nnu-P_comb}, and \ref{fig:dY-nnu_qsos}).  
However, for consistency with the solar system and/or present
interstellar D abundances, such a large primordial D abundance
requires very efficient D destruction.  The low baryon density which
corresponds to high-D still leaves room for dark baryons and
reinforces the case for non-baryonic dark matter (see
Fig.~\ref{fig:H-Omega_B}).  However, when combined with the x-ray
cluster data, low $\Omega_{\rm B}$ and high $f_{\rm HG}$ suggest a
very low density universe ($\Omega_{\rm M} \lesssim 0.21$ for $H_0 \ge
50$ km/s/Mpc; see Fig.~\ref{fig:H-Omega_M}).  The conflict between the
upper bound on $\Omega_{\rm M}$ and the evidence for a lower bound
$\Omega_{\rm DYN} \gtrsim 0.2 - 0.3$ argue against high-D and low
$\eta_{10}$ (see Figs.~\ref{fig:H-Omega_M_comb} and
\ref{fig:H-Omega_ML_comb}) .  Such a low value for $\Omega_{\rm M}$
is also in conflict with the constraint from the shape parameter
$\Gamma$ (see Figs.~\ref{fig:H-Omega_M_comb} and
\ref{fig:H-Omega_ML_comb}).  These problems persist even allowing
for a non-vanishing cosmological constant (which could resolve the
age-expansion rate problem).  See Figs.~\ref{fig:H-Omega_M_comb} and
\ref{fig:H-Omega_ML_comb}.

In contrast, the low-D case leads to severe tension between the SBBN
prediction and the inferred primordial abundance of $^4$He (see
Figs.~\ref{fig:bbn_comb_qsos},
\ref{fig:eta-Y_qsos}, and \ref{fig:eta-y7_qsos}).  This stress on SBBN can
be relieved if the primordial helium mass fraction, derived from
observations of low metallicity HII regions, is in error --- due,
perhaps, to unaccounted systematic effects --- by an amount $\Delta
Y_{\rm sys} \ge 0.011$ (see Fig.~\ref{fig:dY-nnu_qsos}).

Alternatively, this conflict could be evidence of ``new physics''
\cite{Langacker-TASI} ($N_\nu \ne 3$; see Figs.~\ref{fig:nnu-P_comb} 
and \ref{fig:dY-nnu_qsos}).  The best fit between predictions and
observations with low D is for $N_\nu = 1.9 \pm 0.3$.  One way
to alter standard BBN is to change the physics of the neutrino sector.
For example, many models predict the existence of sterile neutrinos,
which interact only by mixing with the ordinary neutrinos.  Such
sterile neutrinos would not contribute significantly to the number of
effective neutrinos ($2.991 \pm 0.016$) inferred from the Z line-shape
\cite{LEP-Nnu-limit}, but could be produced cosmologically for
a wide range of masses and mixings \cite{Langacker-nu_s}.  However,
they only increase $N_\nu$, exacerbating the discrepancy.

Another possibility arises if $\nu_\tau$ has a mass in the range $10 \
{\rm MeV}\lesssim M_{\nu_\tau} \le 24\ {\rm MeV}$ (the upper limit is
the recent result from ALEPH \cite{ALEPH}).  In this case BBN
production of $^4$He can be either increased or decreased (relative to
the standard case), depending on whether $\nu_\tau$ is stable or
unstable on nucleosynthesis time scales ($\sim 1$ sec).  An
effectively stable $\nu_\tau$ ($\tau \ge 10$ sec) in this mass range
always increases $Y$ relative to the standard case \cite{nu-tau} (but,
see \cite{Hannestad-Madsen}) and would thus make for a worse fit with
the data.  However, if $\nu_\tau$ has a lifetime $\lesssim 10$ sec and
decays into $\nu_\mu +\phi$ (where $\phi$ is a `majoron-like' scalar),
\footnote{%
Decays with $\nu_e$ in the final state can directly alter the
neutron-to-proton ratio and thus affect $Y_{\rm P}$ somewhat
differently \cite{Gyuk-Turner}.}  it is possible to decrease the
predicted $Y$ relative to the standard case (see figures 3 and 7 of
Ref.~\cite{Kawasaki-etal}).  Such an unstable $\nu_\tau$ contributes
less than a massless neutrino species at the epoch of BBN, thereby
reducing the yield of $^4$He.  For example, a $\nu_\tau$ with mass $20
- 30$ MeV which decays with a lifetime of $\sim 0.1$ sec reduces
$N_\nu$ by $\sim 0.5 - 1$ (and $Y$ by $\sim 0.006 - 0.013$,
respectively), thus helping to resolve the apparent conflict between
theory and observation.  It is also possible to alter the yield of BBN
$^4$He by allowing $\nu_e$ to be degenerate
\cite{degeneracy}.  If there are more $\nu_e$ than $\bar{\nu}_e$, 
$Y$ is reduced relative to the standard (no degeneracy) case as the
extra $\nu_e$'s drive the neutron-to-proton ratio to smaller values at
freeze-out.  A reduction of $Y$ of $\sim 0.01$ can be accomplished with
a $\nu_e$ chemical potential of $\mu_e/T_\nu \sim 0.03$, corresponding
to a net lepton-to-photon ratio of 0.005.  This is to be compared to
the net baryon asymmetry which is smaller by $\sim$7 orders of
magnitude.  Nevertheless, scenarios for a large lepton asymmetry are
possible \cite{Langacker-Segre-Soni}.  Lastly, one can relax the
assumption that baryons are homogeneously distributed.  However,
inhomogeneous BBN typically results in higher $Y_{\rm P}$, and
therefore does not naturally resolve the $^4$He-D conflict
\cite{inhomogeneous-BBN}.

Provided that the high-$Y$, low-D challenge can be resolved (by
$\Delta Y_{\rm sys} \ge 0.011$ and/or $N_\nu < 3$), low-D is
consistent with the Pop II $^7$Li abundance if there has been a modest
amount of lithium destruction/dilution in the oldest stars (see
Figs.~\ref{fig:bbn_comb_qsos} and \ref{fig:eta-y7_qsos}).  The higher
baryon density for low-D strengthens the case for dark baryons (see
Fig.~\ref{fig:H-Omega_B}), although that for non-baryonic dark matter,
while still very strong, is somewhat weakened.  When folded with the
hot gas bound on the x-ray cluster baryon fraction, a ``cluster baryon
crisis'' persists, arguing for $\Omega_{\rm M} < 1$ (see
Figs.~\ref{fig:H-Omega_M}--\ref{fig:H-Omega_ML_comb}).

\acknowledgements

It is pleasure to thank R.\ Carswell, A.\ Evrard, J.\ Felten, C.\
Hogan, M.\ Persic, P.\ Salucci, R.\ Schaefer, R.\ Scherrer, D.\
Thomas, T.\ Walker, and D.\ Weinberg for useful discussions.  This
work is supported by the Department of Energy Contract No.\
DE-AC02-76-ER01545 at Ohio State University and DE-AC02-76-ERO-3071 at
the University of Pennsylvania.

%
%

\renewcommand{\baselinestretch}{1.3}



\vspace{7ex}
\begin{table}[hbt]
\caption{
%
%
%
The constraints on $\eta_{10}$, $\Omega_{\rm B}h^2$, $N_\nu$, $^4$He
mass fraction ($Y$), and $^7$Li abundance from the high
\protect\cite{Rugers-Hogan} and the combined ISM and 
low QSO-deuterium abundances 
\protect\cite{Tytler-Fan-Burles,Burles-Tytler,Geiss}
along with those from solar system D and $^3$He abundances
\protect\cite{D-paper}.  The errors are for 68\% C.L., while the 
ranges in the parentheses are for 95\% C.L.  }
\label{tab:constraints}
\vspace{1.0ex}
\begin{tabular}{l  c c c}
%
		& High D$_{\rm QSO}$ 
                & Low D$_{\rm QSO}$ \& D$_{\rm ISM}$ 
                & D$_{\odot}$, $^3$He$_{\odot}$ \& BBN\\
\hline\\[-2.5ex]
Obs.\ D/H (10$^{-5}$) 
                & $19 \pm 4$  
                & $2.4 \pm 0.3 \pm 0.3$, $\ge 1.6 \pm 0.2$ 
                & $2.6 \pm 0.9$ \\
\hline
$\eta_{10}$     & $1.8 \pm 0.3$ (1.3 -- 2.7)   
                & $6.4^{+0.9}_{-0.7}$ (5.1 -- 8.2)
                & $5.0 ^{+1.5} _{-0.7}$ (3.5 -- 7.9) \\
$\Omega_{\rm B}h^2$
                & $0.007 \pm 0.001$ 
                & $0.023 \pm 0.003$ 
                & $0.018^{+0.005}_{-0.003}$    \\
                & (0.005 -- 0.010)
                & (0.019 -- 0.030)
		& (0.013 -- 0.029) \\
$N_\nu$         & $2.9 \pm 0.3$ ($\le 3.6$) 
                & $1.9 \pm 0.3$ ($\le 2.4$) 
                & $2.1 \pm 0.3$ ($\le 2.6$) \\
$Y$             & $0.234 \pm 0.002$ 
                & $0.249 \pm 0.001$
                & $0.247^{+0.003}_{-0.002}$ \\
		& (0.231 -- 0.239) 
                & (0.246 -- 0.252)
                & (0.243 -- 0.251) \\
$^7$Li/H (10$^{-10})$
                & $1.5 \pm 0.6$ (0.7 -- 3.0) 
                & $4.7 \pm 0.7$ (3.0 -- 7.8)    
                & $2.9^{+2.0}_{-0.8}$ (1.4 -- 7.3)    \\
%
%
\end{tabular}
\end{table}

\onecolumn
\renewcommand{\baselinestretch}{1.0}
\newlength{\yysize}
\setlength{\yysize}{0.90\hsize}
\newlength{\squaresize}
\setlength{\squaresize}{1.00\hsize}
\newlength{\pslhfsize}
\setlength{\pslhfsize}{1.00\hsize}

%
%

\clearpage

%
%
\begin{figure}[h]

\vspace*{\headroom}
\postscript{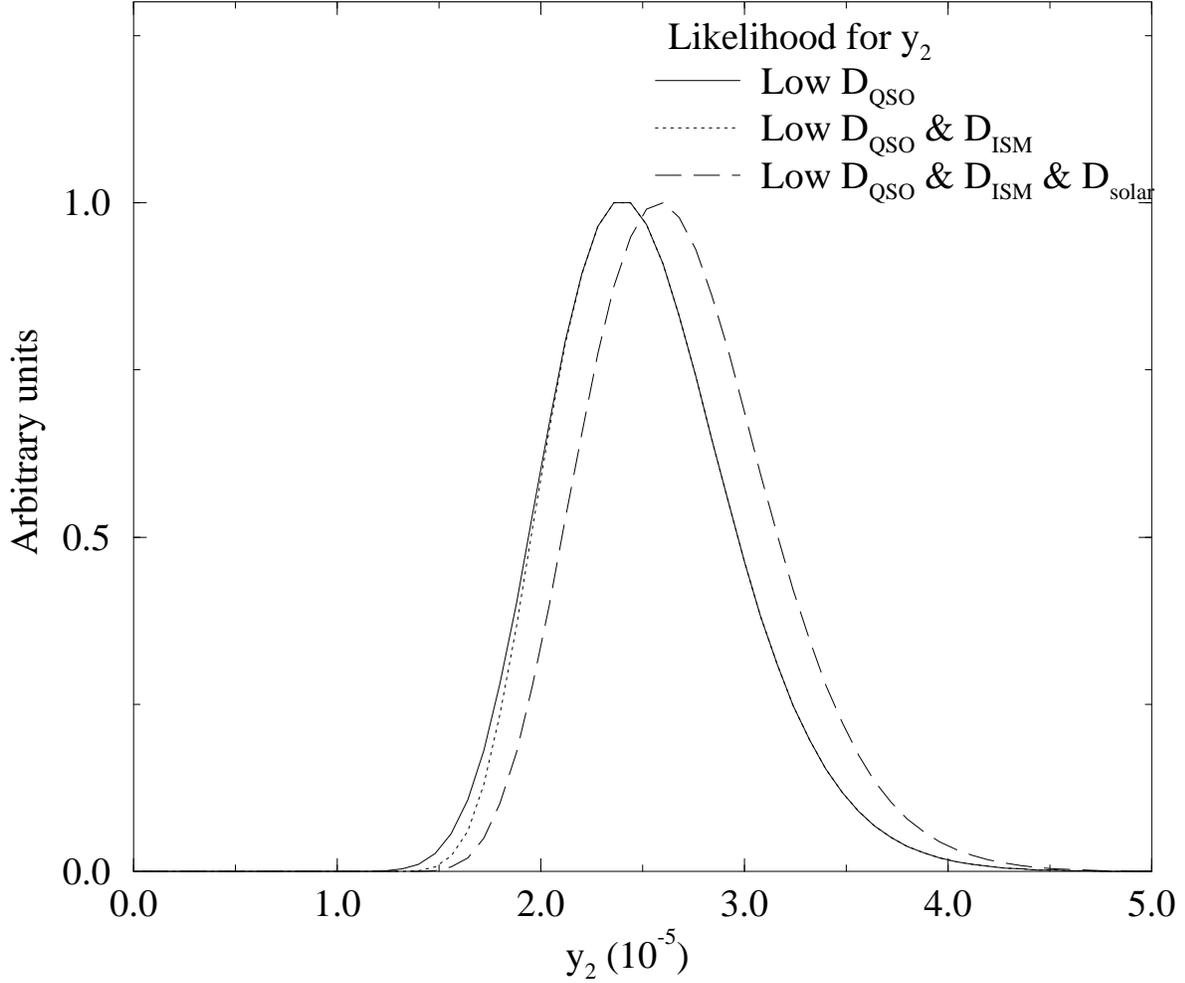}{\pslhfsize}
\vspace{\psfigskip}

\caption{
The likelihood distributions for $y_2$ implied by the low (QSO) D
value (solid line) \protect\cite{Tytler-Fan-Burles}, the combined low
D QSO and ISM limit (dotted line), and the combined limits from low D
and ISM and solar system lower bounds (dashed line).}
\label{fig:y2-P_comb}
\end{figure}

\clearpage

%
%
\begin{figure}[h]

\vspace*{\headroom}
\postscript{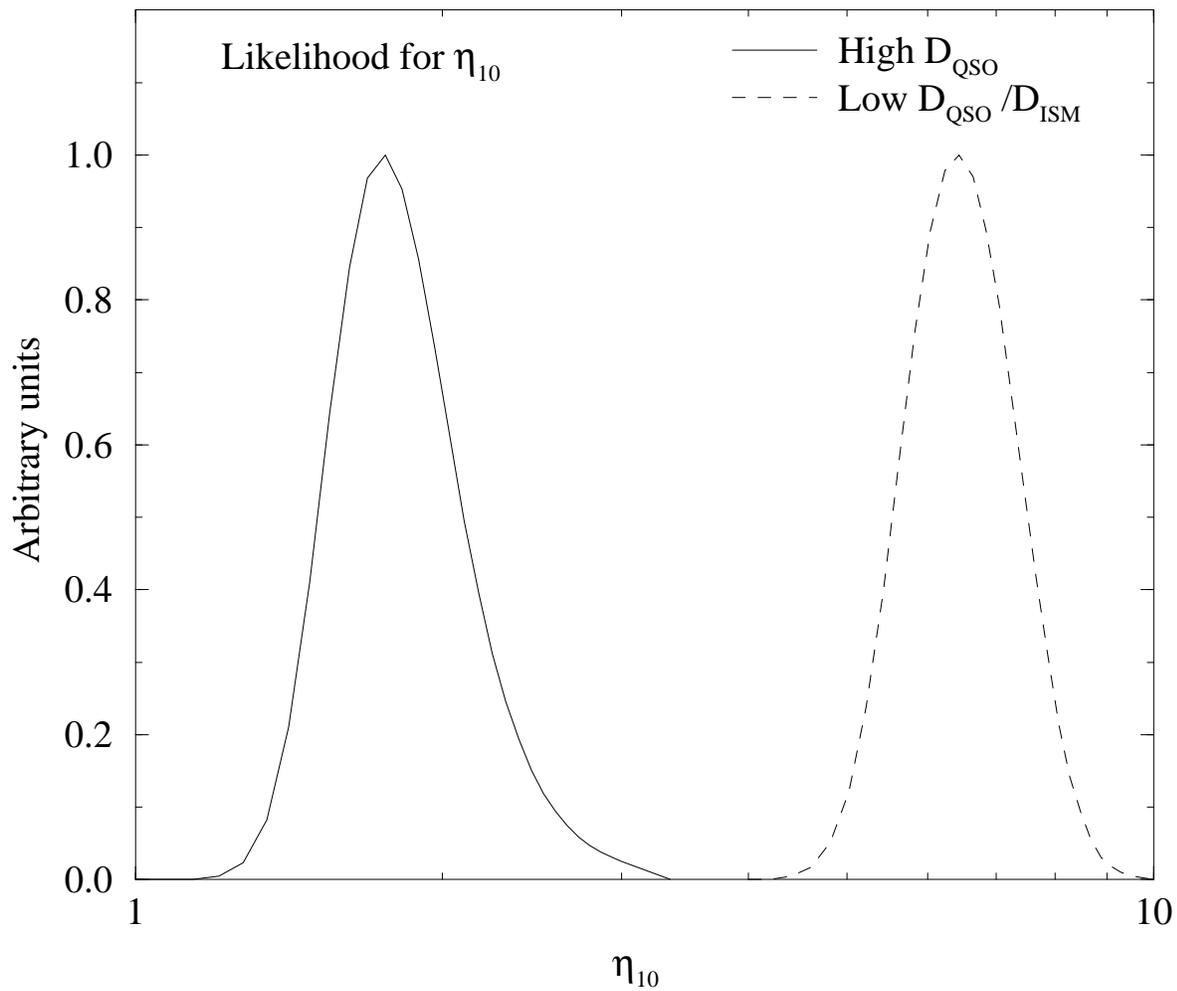}{\pslhfsize}
\vspace{\psfigskip}

\caption{
The likelihood distributions for $\eta_{10}$ implied by the high D
value \protect\cite{Rugers-Hogan} and the low D (plus ISM) value
\protect\cite{Tytler-Fan-Burles,Burles-Tytler}.}
\label{fig:eta-P_qsos}
\end{figure}

\clearpage

%
%
\begin{figure}[h]

\vspace*{\headroom}
\postscript{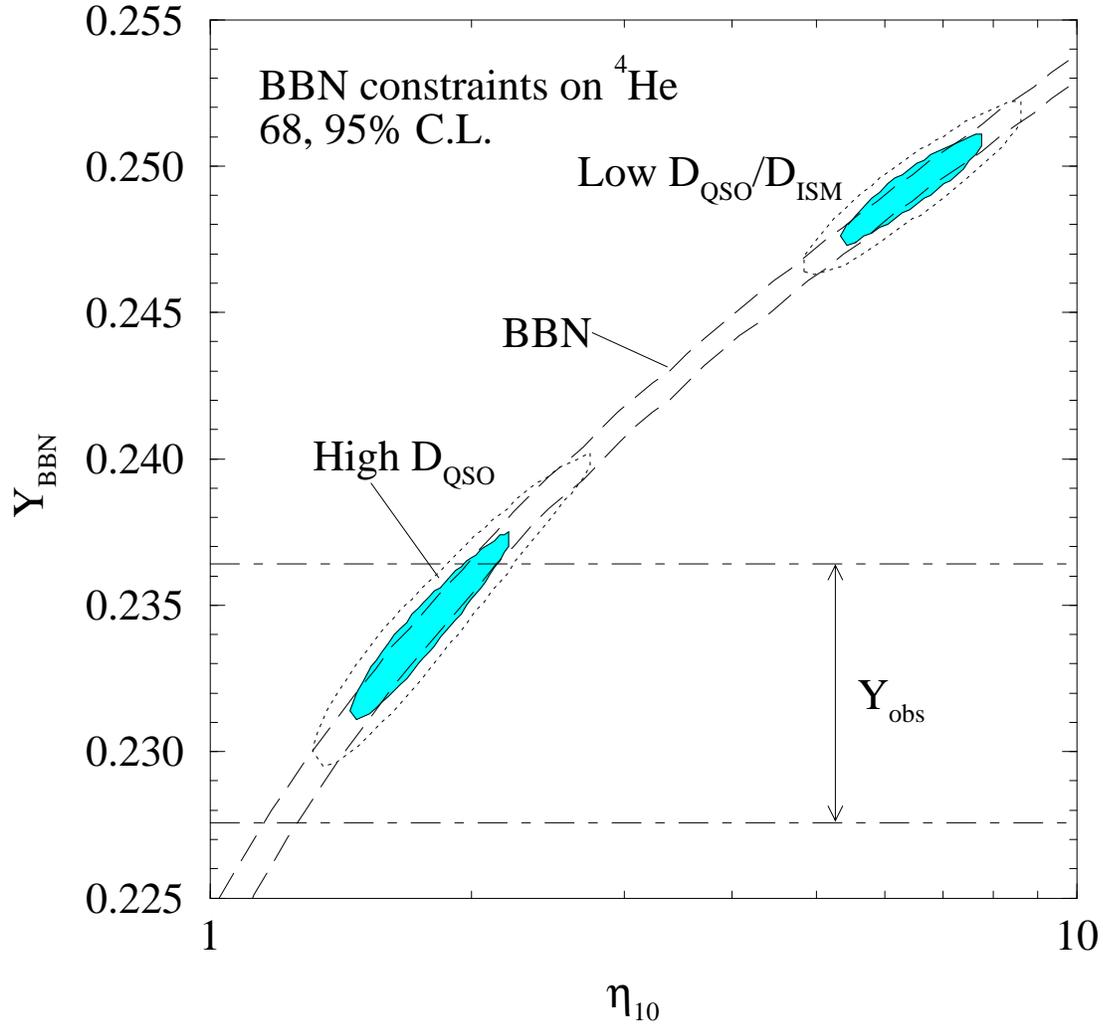}{\squaresize}
\vspace{\psfigskip}

\caption{
The BBN prediction for the primordial $^4$He abundance implied by each
of the two QSO D measurements.  The shaded regions and the dotted
lines correspond to the 68\% and 95\% C.L. constraints.  The $^4$He
abundance derived from HII region observations lies between the
dot-dashed lines (68\% C.L.)  }
\label{fig:eta-Y_qsos}
\end{figure}

\clearpage

%
%
\begin{figure}[h]

\vspace*{\headroom}
\postscript{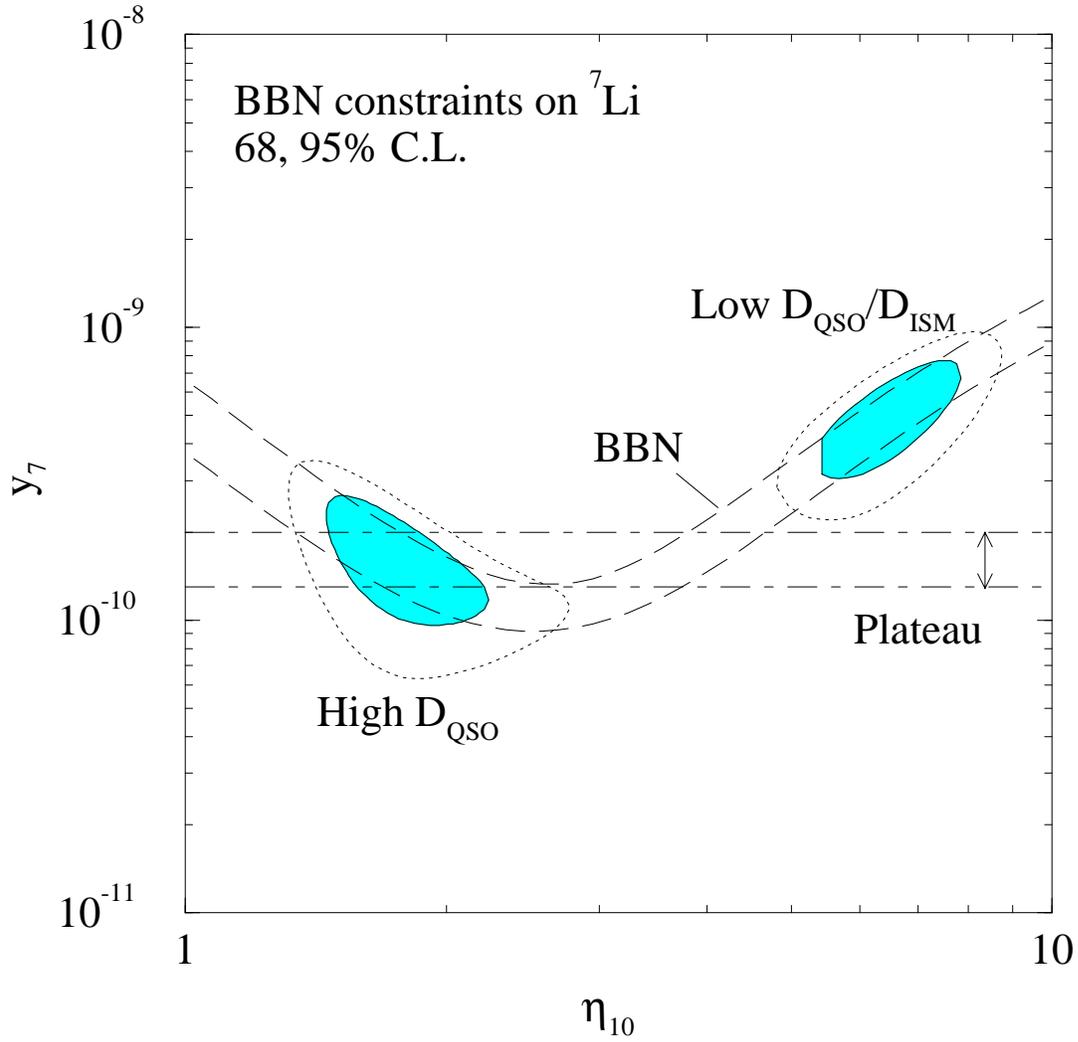}{\squaresize}
\vspace{\psfigskip}

\caption{
The BBN prediction for the primordial $^7$Li abundance implied by each
of the two QSO D measurements.  The shaded regions and the dotted
lines correspond to the 68\% and 95\% C.L. constraints.  The plateau
range (see text) derived from the stars in the galactic halo (ignoring
depletions and creations) is indicated between the dot-dashed lines.}
\label{fig:eta-y7_qsos}
\end{figure}

\clearpage

%
%
\begin{figure}[h]

\postscript{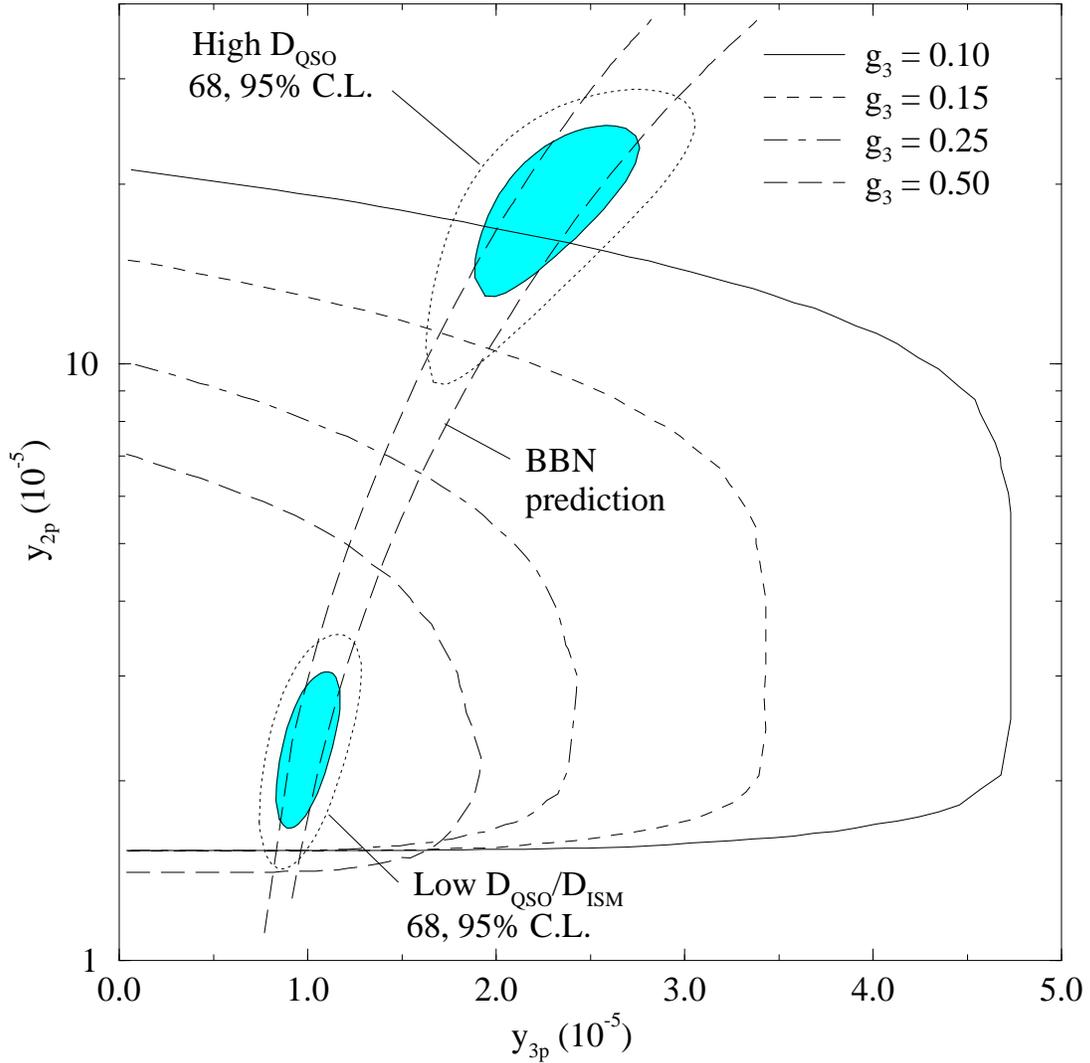}{\yysize}
\vspace{\psfigskip}

\caption{
The two QSO deuterium constraints, combined with the BBN prediction
(long-dashed lines, 1$\sigma$), are shown in the $y_{2P}$--$y_{3P}$
plane (shaded regions at 68\% C.L. and dotted lines at 95\% C.L.).
The regions inside the solid, dashed, dot-dashed, and long-dashed
curves are the abundances consistent with the solar system data for
$g_3$ = 0.10, 0.15, 0.25, and 0.50, respectively
\protect\cite{D-paper}. }
\label{fig:y3-y2_qsos}
\end{figure}

\clearpage

%
%
\begin{figure}[h]

\vspace*{\headroom}
\postscript{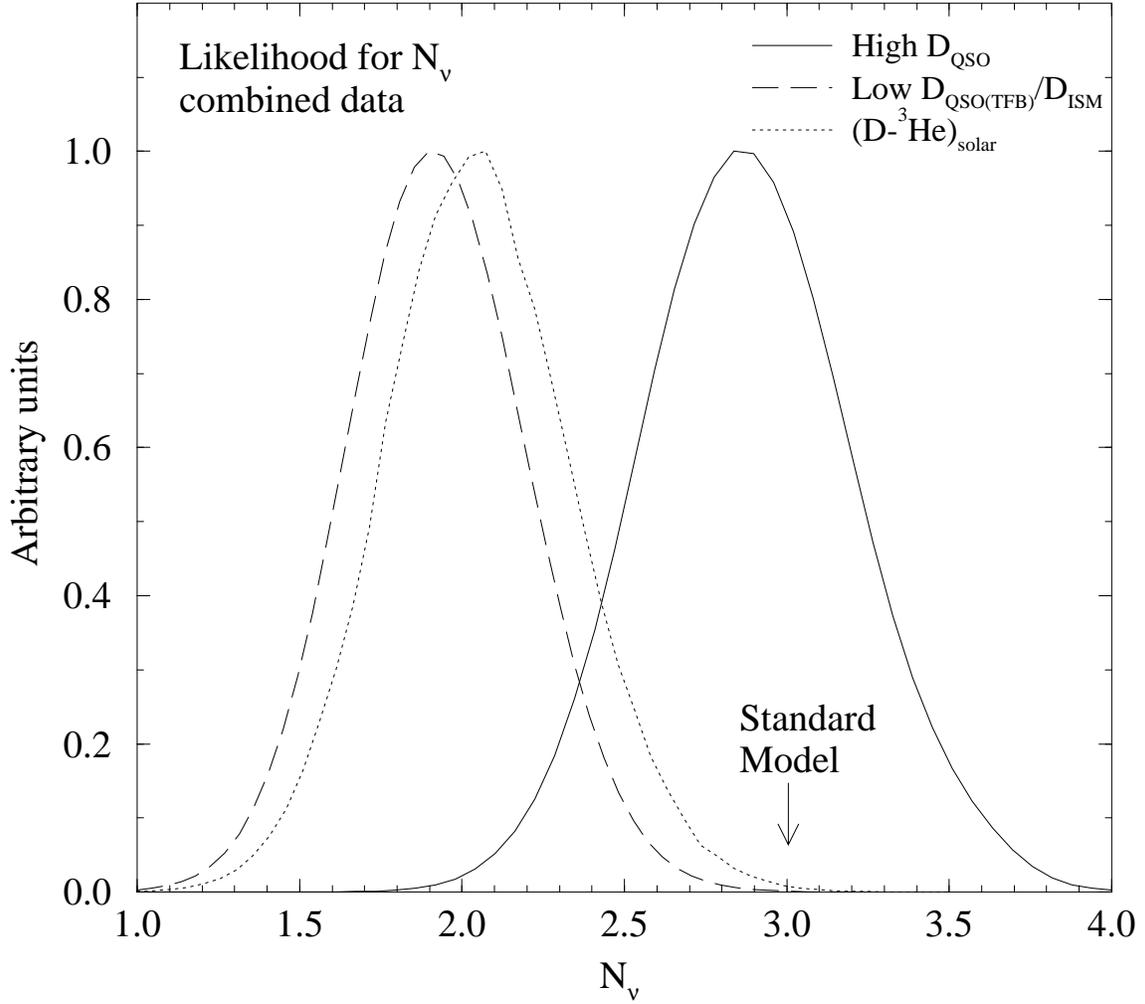}{\pslhfsize}
\vspace{\psfigskip}

\caption{
The likelihood functions for N$_\nu$ derived from the combined
observations of D, $^4$He, and $^7$Li.  The solid, dashed, and dotted
curves are with high D, low D (QSO/ISM), and the solar D and
$^3$He \protect\cite{D-paper}, respectively.  }
\label{fig:nnu-P_comb}
\end{figure}

\clearpage

%
%
\begin{figure}[h]

\vspace*{\headroom}
\postscript{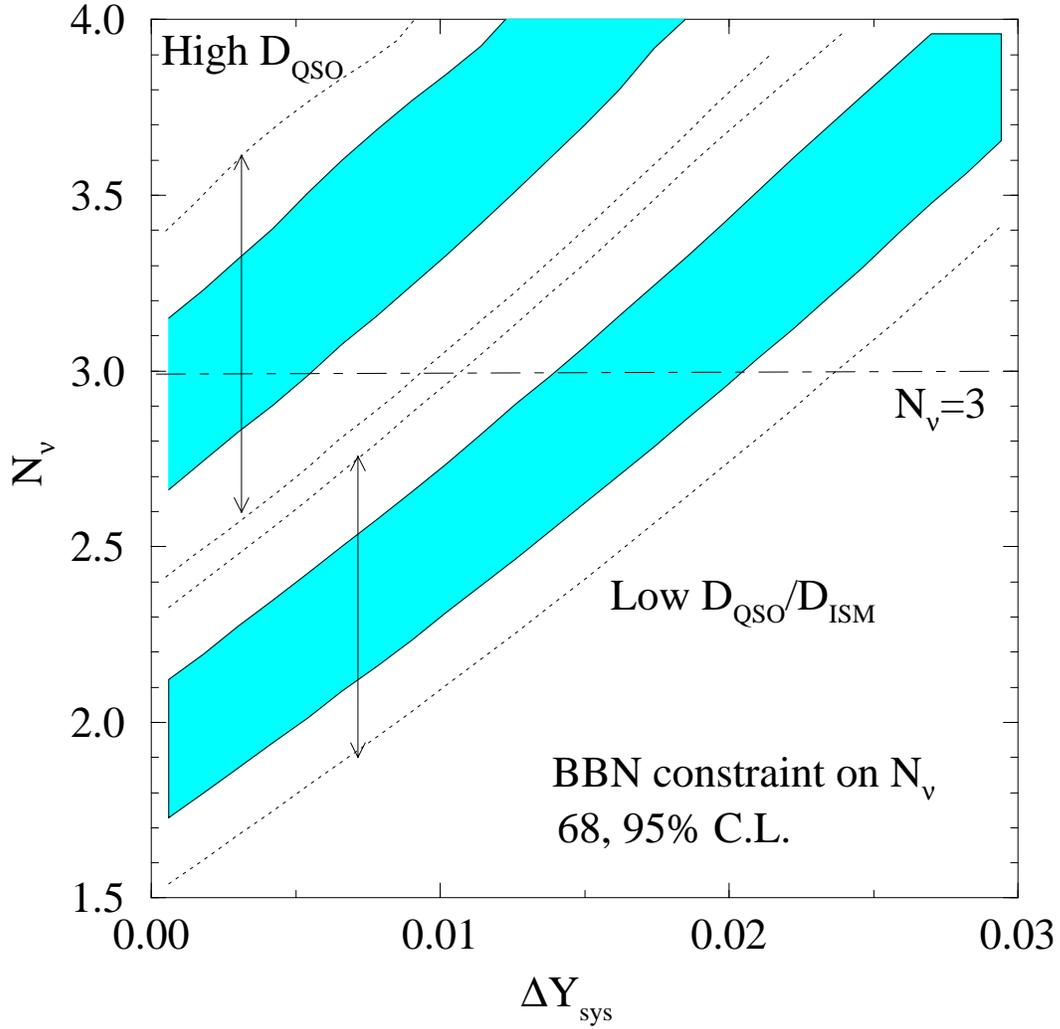}{\squaresize}
\vspace{\psfigskip}

\caption{
The allowed range of $N_\nu$ for high-D and low-D (QSO/ISM) as a
function of systematic offsets ($\Delta Y_{\rm sys}$) in the $^4$He
abundance derived from HII region data.  The shaded regions (dotted
lines) are for 68 (95)\% C.L. }
\label{fig:dY-nnu_qsos}
\end{figure}

\clearpage

%
%
\begin{figure}[h]

\vspace*{\headroom}
\postscript{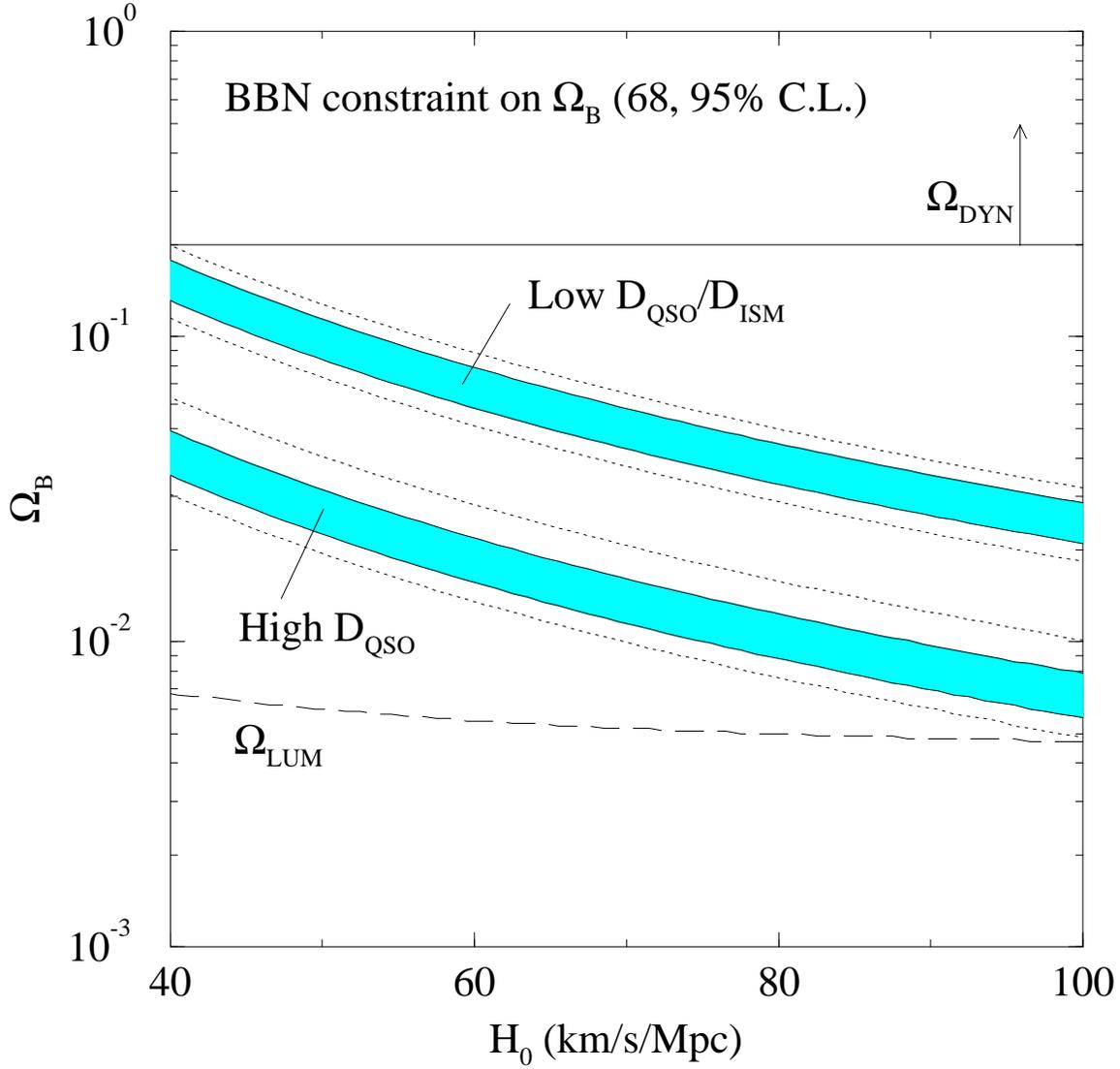}{\squaresize}
\vspace{\psfigskip}

\caption{
The baryon density parameter $\Omega_{\rm B}$ versus the Hubble
parameter $H_{0}$. The two bands correspond to the 68\% (shaded) and
95\% (dotted) C.L. ranges for $\eta_{10}$ inferred from SBBN for high
D$_{\rm QSO}$ and low D$_{\rm QSO}$/D$_{\rm ISM}$ (see Table I).  Also
shown are the estimates \protect\cite{Persic-Salucci} of the
contributions to $\Omega_{\rm B}$ from luminous baryons in galaxies
(dashed curve) and a dynamical estimate
\protect\cite{Ostriker-Steinhardt} of the lower bound to the total
mass density parameter (solid line). }
\label{fig:H-Omega_B}
\end{figure}

\clearpage

%
%
\begin{figure}[h]

\vspace*{\headroom}
\postscript{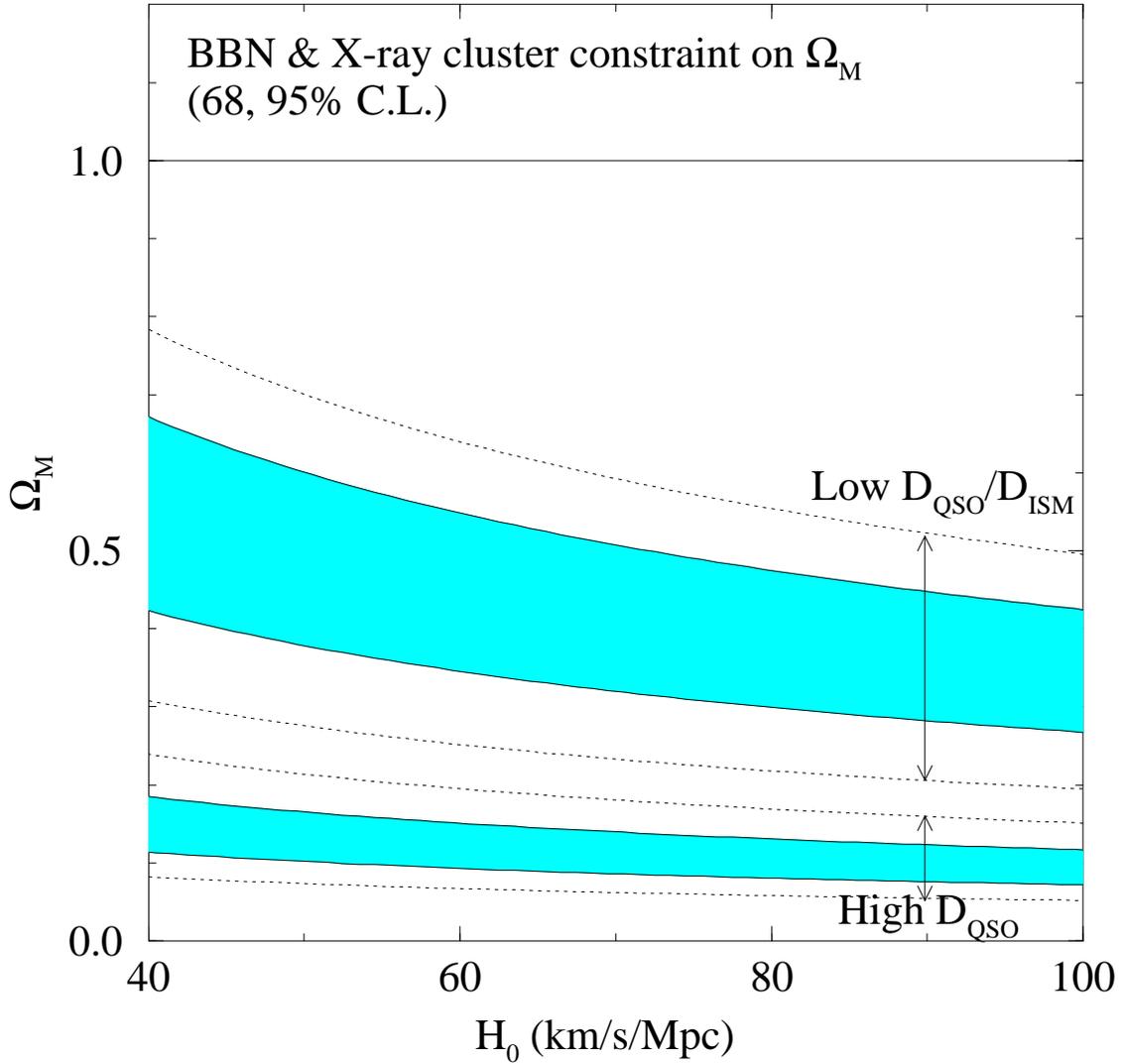}{\squaresize}
\vspace{\psfigskip}

\caption{
BBN and x-ray cluster constraints on the total (matter) mass density
parameter ($\Omega_{\rm M}$) versus the Hubble parameter ($H_{0}$) for
the two choices of primordial D.  The shaded bands (dotted curves) are
the 68\% (95\%) C.L. allowed regions (upper limits; see text).  }
%
\label{fig:H-Omega_M}
\end{figure}

\clearpage

%
%
\begin{figure}[h]

\vspace*{\headroom}
\postscript{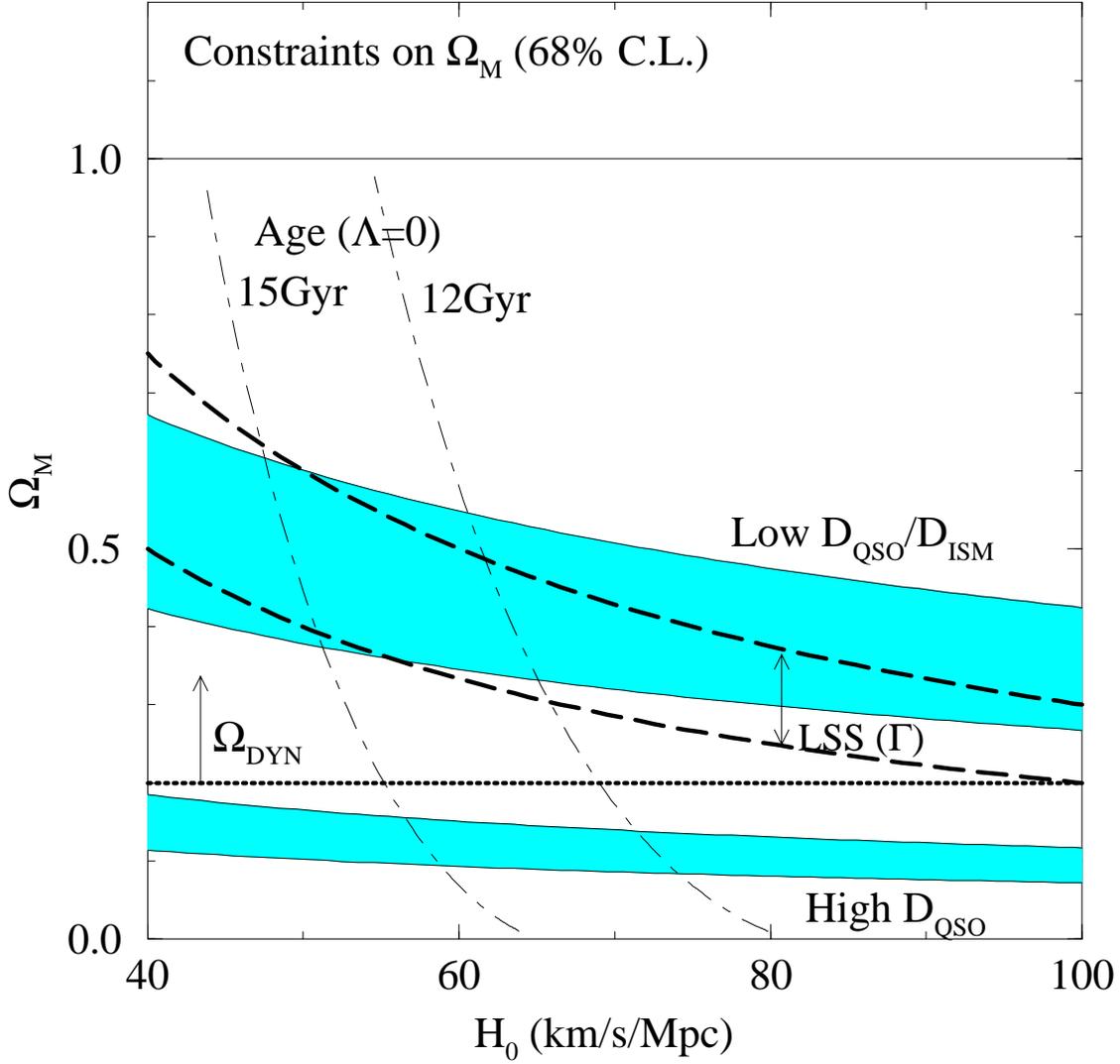}{\squaresize}
\vspace{\psfigskip}

\caption{
The x-ray cluster constraints on the total mass density parameter
($\Omega_{\rm M}$) versus the Hubble parameter ($H_0$) for high
and low D (shaded bands, 68\% C.L.) along with the constraints
 from the shape parameter $\Gamma = \Omega_{\rm M}h$ and the lower
bound to $\Omega_{\rm M}$ from dynamics ($\Omega_{\rm DYN}$).
Also shown are the $\Omega_{\rm M}$ vs.\ $H_0$ relations for two
choices of the present age of the universe.   }
\label{fig:H-Omega_M_comb}
\end{figure}

\clearpage

%
%
\begin{figure}[h]

\vspace*{\headroom}
\postscript{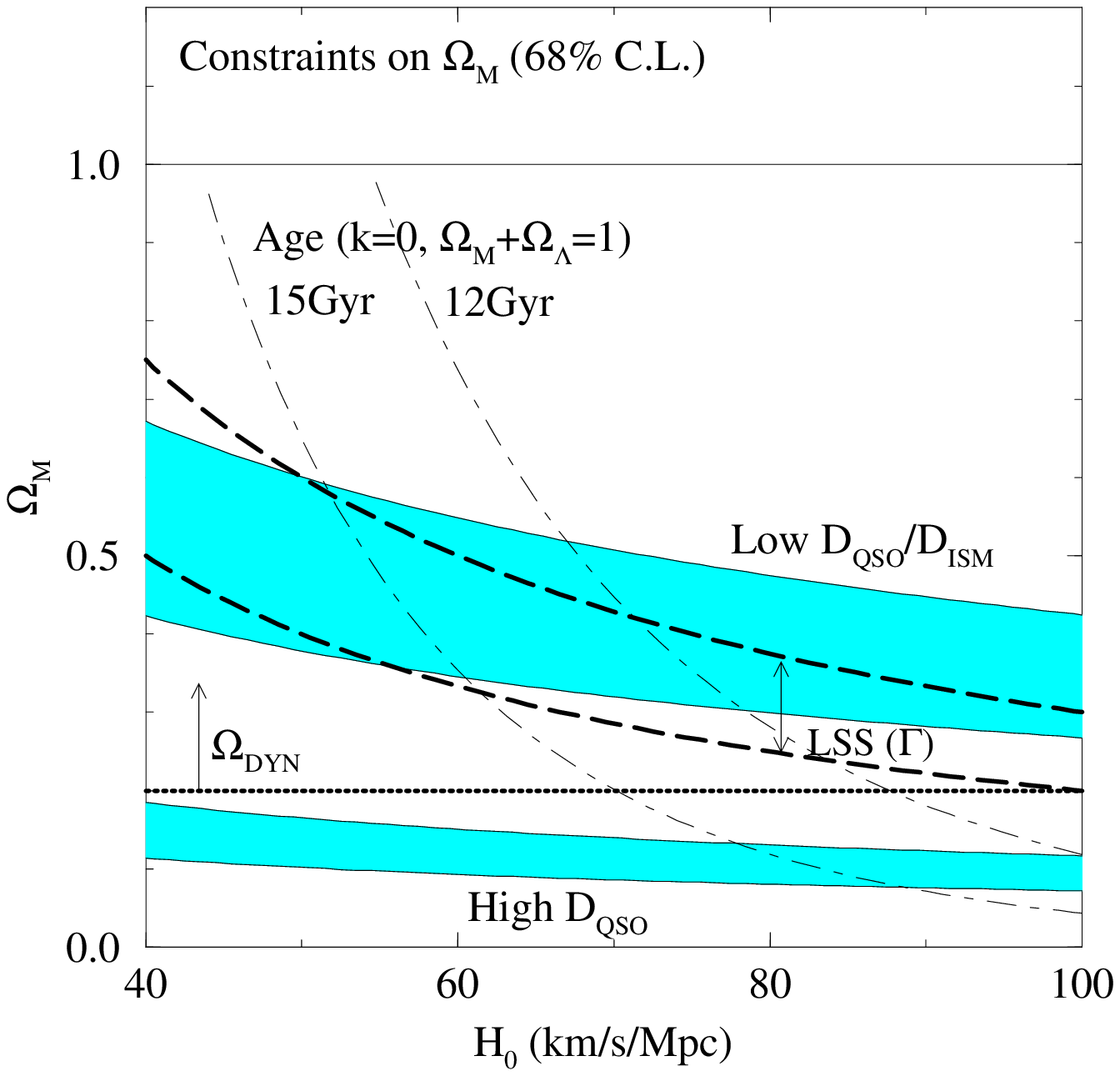}{\squaresize}
\vspace{\psfigskip}

\caption{
As in Fig.~\protect\ref{fig:H-Omega_M_comb}, but for a zero curvature
($k = 0$) model with $\Lambda \ne 0$ ($\Omega_{\rm M} + \Omega_\Lambda
= 1$). }
\label{fig:H-Omega_ML_comb}
\end{figure}

\clearpage


\begin{thebibliography}{99}


\bibitem{Epstein-Lattimer-Schramm}
R.\ Epstein, J.\ Lattimer, and D.\ N.\ Schramm,
Nature {\bf 263}, 198 (1976).

\bibitem{Geiss}
J.\ Geiss,
in {\it Origin and Evolution of the Elements},
edited by N.\ Prantzos, E.\ Vangioni-Flam, and M.\ Casse
(Cambridge University Press, Cambridge, 1993), p.\ 89.

\bibitem{Steigman-Tosi-95}
G.\ Steigman and M.\ Tosi,
Astrophys.\ J.\ {\bf 453}, 173 (1995).

\bibitem{McCullegh}
P.\ R.\ McCullough, 
Astrophys.\ J.\ {\bf 390}, 213 (1992).

\bibitem{Linsky-etal}
J.\ L.\ Linsky {\it et al}.,
Astrophys.\ J.\ {\bf 402}, 694 (1993).

\bibitem{Vangioni-Flam-Audouze}
E.\ Vangioni-Flam and J.\ Audouze,
Astron.\ Astrophys. {\bf 193}, 81 (1988).

\bibitem{Tosi}
M.\ Tosi, 
in {\it From Stars to Galaxies},
edited by C.\ Leitherer, U.\ Fritze von Alvensleben, and J.\ Huchra
(ASP Conference series, 1996).

\bibitem{Steigman-Tosi-92}
G.\ Steigman and M.\ Tosi,
Astrophys.\ J.\ {\bf 401}, 150 (1992).

\bibitem{D-paper}
N.\ Hata, R.\ J.\ Scherrer, G.\ Steigman, D.\ Thomas, and T.\ P.\ Walker,
Astrophys.\ J.\ {\bf 458}, 637 (1996).

\bibitem{Carswell-etal}
R.\ F.\ Carswell, R.\ J.\ Weymann, A.\ J.\ Cooke, and J.\ K.\ Webb,
Mon.\ Not.\ R.\ Astron.\ Soc.\ {\bf 268}, L1 (1994).

\bibitem{Songaila-etal}
A.\ Songaila, L.\ L.\ Cowie, C.\ Hogan, and M.\ Rugers,
Nature {\bf 368} 599 (1994).

\bibitem{Rugers-Hogan}
M.\ Rugers and C.\ J.\ Hogan,
Astrophys.\ J.\ Lett.\ {\bf 459}, 1 (1996).

\bibitem{Tytler-Fan-Burles}
D.\ Tytler, X.\ M.\ Fan, and S.\ Burles,
Los Alamos e-Print archive, astro-ph/9603069
(submitted to Nature).

\bibitem{Burles-Tytler}
S.\ Burles and D.\ Tytler,
Los Alamos e-Print archive, astro-ph/9603070
(submitted to Science).

\bibitem{Edmunds}
M.\ G.\ Edmunds,
Mon.\ Not.\ R.\ Astron.\ Soc.\ {\bf 270}, L37 (1994).

\bibitem{Steigman-94}
G.\ Steigman,
Mon.\ Not.\ R.\ Astron.\ Soc.\ {\bf 269}, L53 (1994).

\bibitem{Carswell-etal-1996}
R.\ F.\ Carswell {\it et al}., 
Mon.\ Not.\ R.\ Astron.\ Soc.\ {\bf 278}, 518 (1996).

\bibitem{Wampler-etal}
E.\ J.\ Wampler {\it et al}.,
Los Alamos e-Print archive, astro-ph/9512084.

\bibitem{crisis-paper}
N.\ Hata, R.\ J.\ Scherrer, G.\ Steigman, D.\ Thomas, T.\ P.\ Walker,
S.\ Bludman, and P.\ Langacker,
Phys.\ Rev.\ Lett.\ {\bf 75}, 3977 (1995).

\bibitem{SBBN-analysis}
D.\ Thomas, N.\ Hata, R.\ J.\ Scherrer, G.\ Steigman, and T.\ P.\ Walker,
work in progress.

\bibitem{Dar}
See, for example, 
A.\ Dar,
Astrophys.\ J.\ {\bf 449}, 550 (1995).

\bibitem{Spite-Spite}
F.\ Spite and M.\ Spite,
Astronomy and Astrophysics {\bf 115}, 357 (1982).

\bibitem{Thorburn}
J.\ A.\ Thorburn,
Astrophys.\ J.\ {\bf 421}, 318 (1994).

\bibitem{Vauclair-Charbonnel}
S.\ Vauclair and C.\ Charbonnel, 
Astron.\ Astrophys. {\bf 295}, 715 (1995).

\bibitem{Molaro-Primas-Bonifacio}
P.\ Molaro, F.\ Primas, and P.\ Bonifacio,
Astron.\ Astrophys. {\bf 295}, 47 (1995).

\bibitem{YTSSO}
J.\ Yang, M.\ S.\ Turner, G.\ Steigman, D.\ N.\ Schramm, and K.\ Olive,
Astrophys.\ J.\ {\bf 281}, 493 (1984).

\bibitem{WSSOK}
T.\ P.\ Walker, G.\ Steigman, D.\ N.\ Schramm, K.\ A.\ Olive, and H.\ Kang,
Astrophys.\ J.\ {\bf 376}, 51 (1991).

\bibitem{Dearborn-Schramm-Steigman}
D.\ S.\ P.\  Dearborn, D.\ N.\ Schramm, and G.\ Steigman,
Astrophys.\ J.\ {\bf 302}, 35 (1986).

\bibitem{Dearborn-Steigman-Tosi}
D.\ Dearborn, G.\ Steigman, and M.\ Tosi,
Astrophys.\ J. in press (vol 465, July 10, 1996).

\bibitem{Palla-etal}
F.\ Palla, D.\ Galli, and J.\ Silk,
Astrophys.\ J.\ {\bf 451}, 44 (1995). 

\bibitem{Pagel-etal}
B.\ E.\ J.\ Pagel, E.\ A.\ Simpson, R.\ J.\ Terlevich, and M.\ G.\ Edmunds,
Mon.\ Not.\ R.\ Astron.\ Soc.\ {\bf 225}, 325 (1992).

\bibitem{Skillman-Kennicutt}
E.\ D.\ Skillman and R.\ C.\ Kennicutt,
Astrophys.\ J.\ {\bf 411}, 655 (1993).

\bibitem{Olive-Steigman-He4}
K.\ A.\ Olive and G.\ Steigman,
Astrophys.\ J.\ Suppl.\ {\bf 97}, 49 (1995).

\bibitem{Sasselov-Goldwirth}
D.\ Sasselov and D.\ Goldwirth,
Astrophys.\ J.\ Lett. {\bf 444}, L5 (1995).

\bibitem{Copi-Schramm-Turner-I}
C.\ Copi, D.\ N.\ Schramm, and M.\ S.\ Turner,
Science {\bf 267}, 192 (1995). 

\bibitem{Persic-Salucci}
M.\ Persic and P.\ Salucci,
submitted to Mon.\ Not.\ R.\ Soc.\ (1995).

\bibitem{Ostriker-Steinhardt}
J.\ P.\ Ostriker and P.\ J.\ Steinhardt,
Los Alamos e-Print Archive, astro-ph/9505066.

\bibitem{Shaya-Peebles-Tully}
E.\ J.\ Shaya, P.\ J.\ E.\ Peebles, and R.\ B.\ Tully,
Astrophys.\ J.\ {\bf 454}, 15 (1995).

\bibitem{Carlberg-Yee-Ellingson}
R.\ G.\ Carlberg, H.\ K.\ C.\ Yee, and E.\ Ellingson,
Los Alamos e-Print Archive, astro-ph/9512087.

\bibitem{Dekel-Rees}
A.\ Dekel and M.\ J.\ Rees,
Astrophys.\ J.\ Lett.\ {\bf 422}, 1 (1994).

\bibitem{White-Frenk}
S.\ D.\ M.\ White and C.\ S.\ Frenk,
Astrophys.\ J.\ {\bf 379}, 52 (1991).

\bibitem{White-Fabian}
D.\ A.\ White and A.\ Fabian,
Mon.\ Not.\ R.\ Astron.\ Soc. {\bf 273}, 72 (1995).

\bibitem{Evrard-Metzler-Navarro}
A.\ E.\ Evrard, C.\ A.\ Metzler, and J.\ F.\ Navarro,
Los Alamos e-Print Archive, astro-ph/9510058 
(submitted to Astrophys.\ J.)

\bibitem{Gould}
A.\ Gould,
Astrophys.\ J.\ {\bf 455}, 44 (1995).

\bibitem{Steigman-Felten}
G.\ Steigman and J.\ E.\ Felten,
Space Science Reviews {\bf 74}, 245 (1995).

\bibitem{Evrard}
A.\ E.\ Evrard,
private communication (1995).

\bibitem{Peacock-Dodds}
J.\ A.\ Peacock and S.\ J.\ Dodds,
Mon.\ No.\ R. Astr.\ Soc.\ {\bf 267}, 1020 (1994).

\bibitem{Liddle-etal}
A.\ R.\ Liddle, D.\ H.\ Lyth, R.\ K.\ Schaefer, Q.\ Shafi, and
P.\ T.\ P.\ Viana,
Los Alamos e-Print Archive, astro-ph/9511057.

\bibitem{Klypin-Nolthenius-Primack}
A.\ Klypin, R.\ Nolthenius, and J.\ Primack,
Los Alamos e-Print Archive, astro-ph/9502062.

\bibitem{Ma-Bertschinger}
C.\ P.\ Ma and E.\ Bertschinger,
Astrophys.\ J.\ Lett.\ {\bf 434}, 5 (1994).

\bibitem{Strickland-Schramm}
R.\ W.\ Strickland and D.\ N.\ Schramm,
Los Alamos e-Print Archive, astro-ph/9511111.

\bibitem{Kofman-etal}
L.\ Kofman, A.\ Klypin, D.\  Pogosian, and J.\ P.\  Henry,
Los Alamos e-Print Archive, astro-ph/9509145.

\bibitem{Langacker-TASI}
See, for example, 
P.\ Langacker,
in {\it Testing the Standard Model (Proceedings of the 1990 Theoretical
Advanced Study Institute in Elementary Particle Physics)},
edited by M.\ Cveti\v{c} and P.\ Langacker 
(World Scientific, Singapore, 1991) p.\ 863.

\bibitem{LEP-Nnu-limit}
A.\ Olshevsky, 
invited talk at the 1995 EPS meeting.

\bibitem{Langacker-nu_s}
P.\ Langacker,
University of Pennsylvania Report No.\ 0401T, 1989 (unpublished).
R.\ Barbieri and A.\ Dolgov, 
Nucl.\ Phys.\ B {\bf 349}, 743 (1991);
K.\ Enqvist, K.\ Kainulainen, and J.\ Maalampi,  
Phys.\ Lett.\ B {\bf 249}, 531 (1990);
M.\ J.\ Thomson and B.\ H.\ J.\ McKellar, 
Phys.\ Lett. B {\bf 259}, 113 (1991);
V.\ Barger {\it et al.}, 
Phys.\ Rev.\ D {\bf 43}, 1759 (1991);
X.\ Shi, D.\ Schramm, and B.\ Fields;
Phys.\ Rev.\ D {\bf 48}, 2563 (1993).

\bibitem{ALEPH}
ALEPH collaboration: D.\ Buskulic {\it et al.},
Phys.\ Lett. B {\bf 349}, 585 (1995).

\bibitem{nu-tau}
E.\ W.\ Kolb and R.\ J.\ Scherrer,
Phys.\ Rev.\ D {\bf 25}, 1481 (1982);
E.\ W.\ Kolb, M.\ S.\ Turner, A.\ Chakravorty, and D.\ N.\ Schramm,
Phys.\ Rev.\ Lett.\ {\bf 67}, 533 (1991);
A.\ Dolgov and I.\ Rothstein,
Phys.\ Rev.\ Lett.\ {71}, 476 (1993);
M.\ Kawasaki, P.\ Kernan, H.-S.\ Kang, R.\ J.\ Scherrer, G.\ Steigman,
and T.\ P.\ Walker,
Nucl.\ Phys.\ B {\bf 419}, 105 (1994);
S.\ Dodelson, G.\ Gyuk, and M.\ S.\ Turner,
Phys.\ Rev.\ D {\bf 49}, 5068 (1994).

\bibitem{Hannestad-Madsen}
S.\ Hannestad and J.\ Madsen,
Los Alamos e-Print Archive, hep-ph/9603201.

\bibitem{Gyuk-Turner}
G.\ Gyuk and M.\ S.\ Turner,
Phys.\ Rev.\ D {\bf 50}, 6130 (1994).

\bibitem{Kawasaki-etal}
M.\ Kawasaki {\it et al}.\ in Ref.~\cite{nu-tau}.

\bibitem{degeneracy}
A. Yahil and  G. Beaudet
Astrophys.\ J.\ {\bf 206}, 26 (1976);
R.\ J.\ Scherrer,
Mon.\ No.\ R. Astr.\ Soc.\ {\bf 205}, 683 (1983);
K.\ A.\ Olive, D.\ N.\ Schramm, D.\  Thomas, and T.\ P.\ Walker, 
Phys.\ Lett.\ B {\bf 265}, 239 (1991);
H.-S.\ Kang and G.\ Steigman,
Nucl.\ Phys.\ B {\bf 372}, 494 (1992).

\bibitem{Langacker-Segre-Soni}
P.\ Langacker, G.\ Segre, and S.\ Soni,
Phys.\ Rev.\ D {\bf 26}, 3425 (1982).

\bibitem{inhomogeneous-BBN}
D.\ Thomas, D.\ N.\ Schramm, K.\ A.\ Olive, G.\ J.\ Mathews,
B.\ S.\ Meyer, and B.\ D.\ Fields,
Astrophys.\ J.\ {\bf 430}, 291 (1994); 
N.\ Terasawa, 
private communication (1994).

\end{thebibliography}
\end{document}